\documentclass[twocolumn]{aastex631}

\usepackage{amsmath,amssymb,times}
\usepackage{acronym}

\shorttitle{Implications of Different Surface Flux Transport Models}

\shortauthors{Barnes et al.}

\def\refereeout#1{}

\begin{document}

\title{Implications of Different Solar Photospheric Flux-Transport Models for Global Coronal and Heliospheric Modeling}

\author[0000-0003-3571-8728]{Graham~Barnes} \affil{NorthWest Research
  Associates, 3380 Mitchell Ln., Boulder, CO 80301, USA}

\author[0000-0002-6338-0691]{Marc~L.~DeRosa} \affil{Lockheed Martin Solar and
  Astrophysics Laboratory, 3251 Hanover St. B/252, Palo Alto, CA 94304, USA}

\author[0000-0001-9498-460X]{Shaela~I.~Jones} \affil{NASA Goddard Space Flight Center, Greenbelt, MD USA} \affil{Department of Physics, Catholic University of America, Washington, DC USA}

\author[0000-0001-9326-3448]{Charles.~N.~Arge} \affil{NASA Goddard Space Flight Center, Greenbelt, MD USA}

\author[0000-0002-6038-6369]{Carl~J.~Henney} \affil{Air Force Research Laboratory, Space Vehicles Directorate, Kirtland AFB, NM USA}

\author[0000-0003-2110-9753]{Mark~C.~M.~Cheung} \affil{CSIRO, Space \& Astronomy, PO Box 76, Epping, NSW 1710, Australia}

\begin{abstract}
The concept of surface-flux transport (SFT) is commonly used in evolving models of the large-scale solar surface magnetic field. These photospheric models are used to determine the large-scale structure of the overlying coronal magnetic field, as well as to make predictions about the fields and flows that structure the solar wind. We compare predictions from two SFT models for the solar wind, open magnetic field footpoints, and the presence of coronal magnetic null points throughout various phases of a solar activity cycle, focusing on the months of April in even-numbered years between 2012 and 2020, inclusive. We find that there is a solar cycle dependence to each of the metrics considered, but there is not a single phase of the cycle in which all the metrics indicate good agreement between the models. The metrics also reveal large, transient differences between the models when a new active region is rotating into the assimilation window. The evolution of the surface flux is governed by a combination of large scale flows and comparatively small scale motions associated with convection. Because the latter flows evolve rapidly, there are intervals during which their impact on the surface flux can only be characterized in a statistical sense, thus their impact is modeled by introducing a random evolution that reproduces the typical surface flux evolution. We find that the differences between the predicted properties are dominated by differences in the model assumptions and implementation, rather than selection of a particular realization of the random evolution.
\end{abstract}


\section{Introduction}

The evolution of the Sun’s magnetic field shapes the corona and heliosphere. However, the only place where it is routinely measured over the entire disk of the Sun is the photosphere. Knowledge of the global photosphere magnetic field is important, as it serves as a key driver to virtually all models of the corona. However, only the Earth-facing side of the Sun is routinely observed, necessitating modeling to estimate the instantaneous state of the global photospheric magnetic field.

Synoptic charts, in which a sequence of observations close to central meridian are stacked together to form a global chart as the Sun rotates are the traditional way to construct a global map of the photospheric magnetic field, but this scheme suffers from the fact that some of the observations are close to a month old (due to the Sun's rotation rate of about one month), and results in large discontinuities between old and recent observations. One improvement on this scheme is to account for the evolution of the photospheric flux that is known to occur. Surface-flux transport (SFT) models aim to capture these dynamics, and apply the known evolutionary processes affecting the magnetic field in areas of the solar surface for which no current observations are available, using a combination of large-scale flows (e.g., differential rotation and meridional flow) to advect the surface flux, and an approximation to the effects of the small-scale flows from convection or supergranulations. A number of such SFT models exist \citep[e.g.,][]{schr2001a,UptonHathaway2014,hick2015}, with a common goal of determining the radial magnetic field over the full surface of the Sun at a given moment in time, but the assumptions and implementation differ among these models. 

While SFT models are sometimes used for other purposes, such as the prediction of extreme ultraviolet (EUV) irradiance \citep[e.g.,][]{Henney2015} or upcoming solar cycles \citep[e.g.,][]{Upton2018}, one of their more common uses is as boundary conditions for global coronal and heliospheric models \citep{Jiangetal2014}. Although SFT models typically have the same broad goal of modelling the evolution of the global photospheric magnetic field, differences in implementation can affect the resulting photospheric maps. In this study, we assess these differences by comparing the predictions of  solar wind properties, the location of the footpoints of open magnetic flux, and the presence of coronal magnetic null points based on two different SFT models. Other complementary studies have focused on the impact of different data sources \citep{Rileyetal2014}, the impact of using a global boundary map from a synoptic chart compared with a SFT model \citep{JinNittaCohen2022}, or on comparisons with observations \citep[e.g.,][]{Badmanetal2022,Jonesetal2022}, but here we focus on comparing predictions from two SFT models using the same data source to quantify the impact of the different SFT models on coronal and heliospheric models.

\section{Model Descriptions}

In the comparison exercise described here, two different SFT models are used. The goal of each model is to assimilate photospheric magnetogram data and evolve the global photospheric magnetic field forward in time in a realistic manner. The two models considered are the Air Force Data Assimilative Photospheric Flux Transport (ADAPT) model, and the SFT model distributed with the \texttt{pfss} package available through SolarSoft (SSW-PFSS); for the interval analyzed, the Helioseismic and Magnetic Imager \citep[HMI;][]{Scherreretal2012} on board NASA's Solar Dynamics Observatory \citep[\textit{SDO};][]{PesnellThompsonChamberlin2012} is the source of the magnetogram data. 

\begin{figure}[ht!]
\includegraphics[width=\columnwidth]{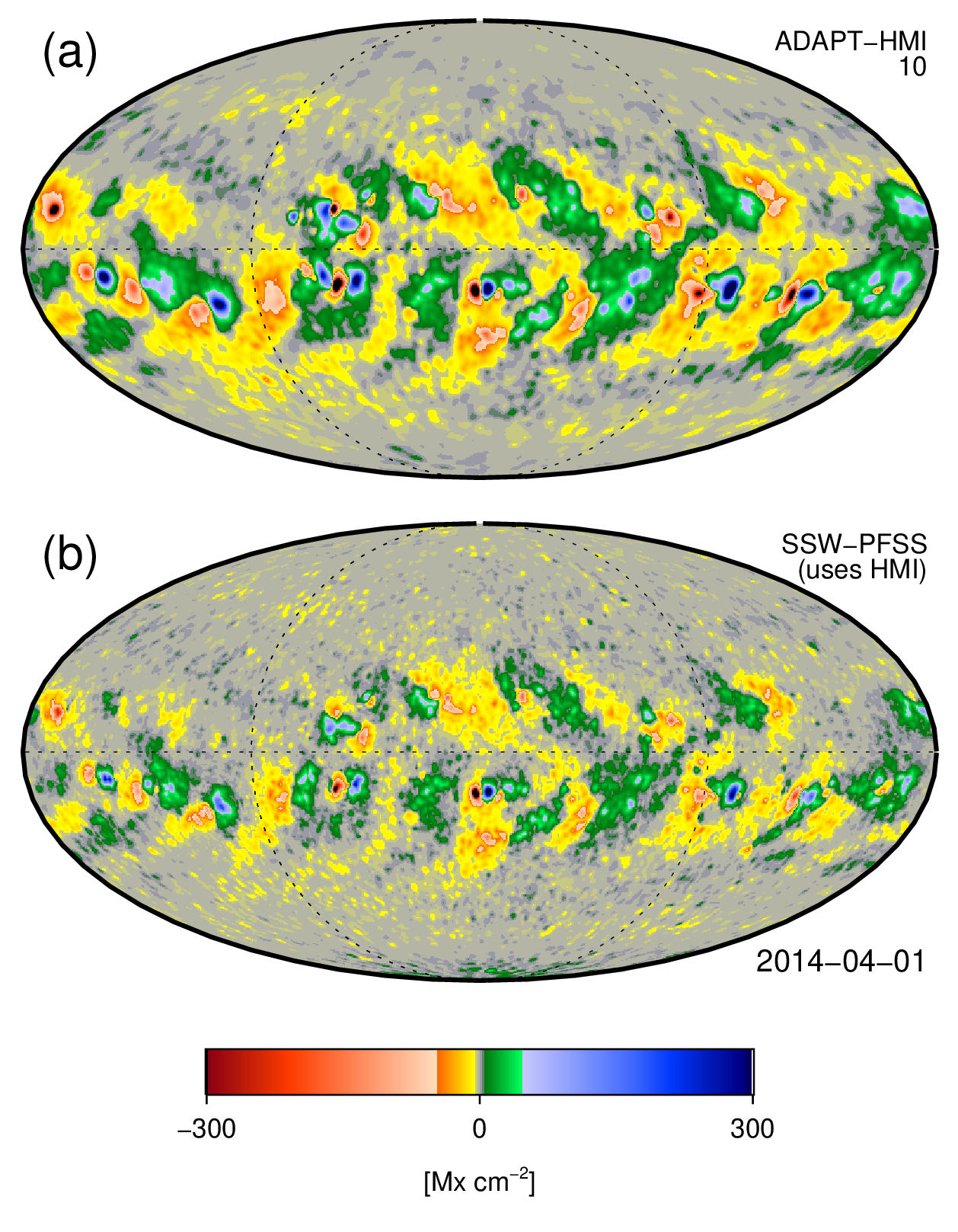}
\caption{Maps of flux density at a radius of 1.005~$R_\odot$ (approximately the base of the corona) for 2014 April~1 for (a) one of the ADAPT-HMI realizations and (b) the SSW-PFSS model. Mollweide map projections are used, with the centroid longitude chosen to be the Carrington longitude as viewed from Earth for this date. In each map, the dashed circle indicates the solar limb as viewed from Earth, and the horizontal dashed line represents the solar equator. The color table has sharp breaks at $\pm$50~Mx~cm$^{-2}$ that allow the structure of both weak and strong magnetic fluxes to be evident. (This color table is courtesy of the SDO/HMI team at Stanford University and can be downloaded from \url{http://jsoc.stanford.edu/jsocwiki/MagneticField}.)
This figure is sampled from an animation (available online) consisting of one image per day for each day in April of even numbered years between 2012 and 2020. For both ADAPT-HMI and SSW-PFSS, the animation shows a similar pattern of active regions being present at lower latitudes as the solar cycle progresses to be almost completely absent near solar minimum, but there are differences in the polar field between the two models, from opposite polarity of the weak polar fields near solar maximum to systematically stronger polar field in ADAPT-HMI near solar minimum.}
\label{fig:map-at-cbase}
\end{figure}

Figure~\ref{fig:map-at-cbase} shows a global magnetic field map provided by each of the models, sampled on 2014 April~1. In this figure and in Figures~\ref{fig:pil-at-ssurf-ADAPT-set}--\ref{fig:openflux-at-lowcorona} that follow, Mollweide map projections are used. This projection was chosen as an equal-area projection that enables the full 4$\pi$ steradians of a spherical surface to be visualized without greatly distorting the shape of features. In the set of figures, the centroid of the map projections is aligned with the Carrington longitude of the central meridian for 2014 April~1. Both models assimilate magnetogram data into the model within $\sim 60^\circ$ of the center of the Earth-facing side of the Sun. As a result, most of the Earth-facing longitudes in the model (which appear within the dashed circle in the map projection) are based on data that are more current, resulting in closer agreement between the models, while polar areas and farside longitudes possess more uncertainty. One result of only observing a third of the photosphere well at any given time is a monopole moment that is frequently non-zero. The monopole residual after data assimilation can be the result of an active region being clipped at the assimilation window or artificial imbalances created with the radial field assumption for line-of-sight observations, especially when the central meridian distance of a region, e.g., is greater than 30 degrees.  In addition, the modeled supergranular flows, along with the differential rotation, for periods of approximately two weeks on the farside will result in misalignment of weak and strong flux regions at the east-limb assimilation boundary. We next describe each model in turn, and provide a more detailed comparison of the imposed large-scale flows in Appendix~\ref{sec:flow-profiles}.

\subsection{ADAPT} \label{sec:ADAPT}

The ADAPT model \citep{arge10,arge13,hick2015,Schonfeld2022} utilizes flux transport, based on the \citet{Worden2000} model, to account for differential rotation, along with meridional and supergranulation flows, when observational data are not available. The diffusion of flux is estimated in ADAPT utilizing random sinks with lifetimes of about 24 hours that approximate stochastic supergranulation flows, based on \cite{Mosher1977} and \cite{SimonTitleWeiss1995}.  In addition, the ADAPT model incorporates new magnetogram input using the ensemble least-squares data assimilation technique to account for both model and data uncertainties when new maps are generated \citep{hick2015}.  For example, ADAPT heavily weights observations taken near the central meridian where magnetograms are most reliable, while the model specification of the field is generally given more weight near the limbs where observations are the least reliable. To limit the observational errors near the limb, that arise primarily from foreshortening and the highly variable horizontal magnetic signal that increases toward the limb \citep[see, e.g.,][]{Harvey2007}, the ADAPT model does not assimilate data beyond about 70 degrees from disk center. 

ADAPT provides an ensemble of possible states (i.e., realizations) of the solar surface magnetic field, which attempts to realistically represent the uncertainly in our knowledge of the global photospheric magnetic flux distribution at any given moment in time.  The realizations differ in the locations of the sinks used to model the effects of supergranulation, so the flow directions are different between realizations and change within a realization on the timescale of a day. Early ADAPT model development testing with various meridional flow profiles resulted in small changes in the polar mean field when compared to different supergranulation flows.  Thus, for this study, the ADAPT ensemble variance is driven purely by different supergranulation flows.

The ADAPT model scales input magnetogram data, boot-strapped from KPVT 1992 magnetograms, to minimize discontinuities in derived products, e.g., F10.7 (solar radio flux at 10.7 cm, i.e., 2.8 GHz) and EUV \citep[see][]{Henney2012,Henney2015}. To date, all ADAPT maps for a given date (i.e., based on VSM, GONG, or HMI input magnetograms) originate, and are a continuation, from a KPVT seed map beginning in 1992. The HMI data are taken from the \texttt{hmi.Mrmap\_latlon\_900x900\_720s} series, which is derived from the standard line-of-sight processing pipeline, and is available through the JSOC (\url{http://jsoc.stanford.edu/ajax/lookdata.html}). In addition, at each assimilation step, the ADAPT model nominally rescales each realization to dampen the monopole moment (i.e., the zero-point offset) residual by 80 percent. Furthermore, the ADAPT maps used here include spatial smoothing (i.e., a spatial convolution with a Hanning window function) to preserve the area resolution at the equator throughout the map. 

\subsection{SSW-PFSS}

The \texttt{pfss} software package, which is distributed via the Solar SoftWare distribution platform (colloquially known as ``SSW'' or ``SolarSoft''; \citealt{free2001}), enables users to access global maps of the photospheric magnetic field and the associated potential field extrapolations dating back to 1996. The photospheric magnetic field maps are based on a SFT model into which full-disk magnetogram data from MDI (before 2010) and HMI (since 2010) are assimilated. Hereafter, we refer to this model as the ``SSW-PFSS'' model.

Unlike the ADAPT model, the SSW-PFSS model tracks individual concentrations of magnetic flux treated as point sources. The evolutionary aspects of this SFT model are described in detail in \citet{schr2001a}, and includes processes that advect flux concentrations across the model photosphere due to time-independent flow profiles for the solar differential rotation and meridional circulation (see Appendix~\ref{sec:flow-profiles}). The model also includes a flux-dependent dispersal scheme that approximates the effect of convection on surface magnetic elements as well as a prescription for the spontaneous fragmenting of flux elements, both of which are based on observations \citep{1996ApJ...468..921S,1997ApJ...487..424S}. Cancellation of flux also occurs in the model. Additionally, there is a flux-decay term that removes flux on solar-cycle timescales, which was found to be necessary in order to run the model for multiple sunspot cycles \citep{schr2002}.

The assimilative process by which magnetograms are incorporated into the model are detailed in \cite{schr2003}. In essence, during assimilation, for all points in the SFT model within approximately 60$^\circ$ of disk center, the modeled flux is replaced by full-disk line-of-sight magnetogram data, corrected from $B_\text{los}$ to $B_r$. The MDI data are the Level~1.8 magnetogram data provided by the instrument team, and the HMI data are taken from the \texttt{hmi.M\_45s} dataseries at the JSOC. Before assimilation into the model, care is taken to remove instrumental artifacts in the magnetogram data, including variations in the determinations of $B_\text{los}$ across the field of view, the $B_\text{los}$ zero point, and a correction factor to match the $B_\text{los}$ in HMI magnetograms to that in MDI \citep{liuy2012}. The model was initialized using a dipole field at the beginning of the SOHO era, after which MDI data were assimilated into the model at a cadence of once every 96~minutes. Following the cutover from MDI to HMI, data from HMI were (and continue to be) assimilated into the SFT model once per hour.

\subsection{Imposing Flux Balance}\label{sec:monopole}

The partial view of the global photospheric magnetic field afforded by SDO's position in Earth orbit results in magnetogram images that are not flux-balanced. This aspect introduces a flux imbalance into the SFT models that must be addressed somehow. These errors, while small in a statistical sense (usually of order a few percent in the ratio of net flux to absolute flux integrated over the global magnetic field map), build up over time and can affect the large-scale structure of the coronal magnetic field and the downstream wind solutions. As a result, we apply a correction scheme to both SFT models that balances the net flux in the magnetic maps while simultaneously preserving the global absolute flux in the models \citep{Jonesetal2020}. The correction scheme also preserves the sign of the flux density at each location, and as a result the contours of polarity inversion lines at the photosphere do not change.

The flux-balancing scheme works as follows.  For a given global magnetic map $B_r$, we first determine the net flux $\Phi_\text{tot}=\oint_S B_r dS$ by integrating $B_r$ over the photospheric surface $S$. Maps that are not flux-balanced will have $\Phi_\text{tot}\ne 0$. Next, we define a mask $M_+$ that is equal to unity at locations where the polarity is positive and zero elsewhere, and a complementary mask $M_-$ that equals unity where the polarity is negative and zero elsewhere. Using the two masks to integrate $B_r$ over $S$ results in the total (global) absolute flux in positive and negative areas of the full-Sun magnetic map, which we denote as $\Phi_+$ and $\Phi_-$:
\begin{equation}\begin{aligned}
    \Phi_+ &= \oint_S M_+ |B_r| dS\\
    \Phi_- &= \oint_S M_- |B_r| dS
    \end{aligned}\end{equation}
The fact that the absolute value of $B_r$ is used in the integrals means that both $\Phi_+$ and $\Phi_-$ are positive-definite quantities. The total (global) absolute flux $\Phi_\text{abs}$ for this model is therefore $\Phi_++\Phi_-$. Any flux imbalance is embodied by $\Phi_\text{tot}=\oint_S B_r dS$, or equivalently $\Phi_\text{tot}=\Phi_+-\Phi_-$.

The flux-balancing scheme used here involves scaling $\Phi_+$ and $\Phi_-$ by different factors. Two scalar quantities $\alpha_+$ and $\alpha_-$ are introduced, such that $\alpha_+$ will be multiplied to all pixels in the map with a positive polarity and $\alpha_-$ will be multiplied to all pixels having a negative polarity. If the net flux of the new magnetic map were to vanish (thereby removing the flux imbalance) while still preserving the total absolute flux, this implies
\begin{equation} \label{eq:fluxbal}
    \begin{aligned}
    \alpha_+\Phi_+ - \alpha_-\Phi_- &= 0\\
    \alpha_+\Phi_+ + \alpha_-\Phi_- &= \Phi_\text{tot}.
    \end{aligned}
\end{equation}
Solving for $\alpha_+$ and $\alpha_-$ yields
\begin{equation}
    \begin{aligned}
    \alpha_+ &= \frac{\Phi_\text{tot}}{2 \Phi_+}\\
    \alpha_- &= \frac{\Phi_\text{tot}}{2 \Phi_-}.
    \end{aligned}
\end{equation}
Once $\alpha_+$ and $\alpha_-$ have been determined, we apply the flux-balancing correction scheme from Equation~(\ref{eq:fluxbal}) to the output of both SFT models.

\section{Points of Comparison} \label{sec:points-comparison}

The SFT models were run forward in time, and sampled once per day during April in even-numbered years between 2012 and 2020, inclusive, which spans much of sunspot cycle 24 \citep{DVN/EFJXHY_2022,DVN/CFF76Z_2022}. The global magnetic maps output by the SFT models were used to initialize the standard Wang-Sheeley-Arge \citep[WSA;][]{ArgePizzo2000,Argeetal2003,Argeetal2004,McGregoretal2008} coronal and solar-wind modelling framework. The end results include PFSS extrapolations between 1$R_\odot$ and 2.5$R_\odot$, and solar wind polarities and speeds at distances up to 1\,AU. From these outputs, several features of the different models are compared: the open-field regions at the photosphere and the locations of heliospheric sector boundaries (\S\ref{sec:openfield}), the speed of the solar wind at $5.0\,R_\odot$ (\S\ref{sec:windspeed}), and the locations of coronal magnetic null points (\S\ref{sec:nullpoints}).

A point at the photosphere is considered to be located within an open field region if a field line originated there reaches the source surface, and a closed field region if it does not. Null points in the volume of the PFSS extrapolation are located using the trilinear method of \cite{HaynesParnell2007}. A clustering algorithm, described in detail in Appendix~\ref{sec:nullmatch}, is used to determine if coronal null points of the same sign in two models are considered to be the same null point.

We use the Jaccard similarity index to quantitatively compare the open-flux regions and the coronal magnetic null points. Generally, the Jaccard index is a metric that compares two populations that may belong to either (or both) of two sets, and is defined as the number of elements belonging to both sets divided by the total number of elements. Stated more succinctly, the Jaccard index is the ratio of the area of intersection to the area of the union. The Jaccard index is between~0 and~1 inclusive, with indices near~1 indicating that most elements belong to both sets (i.e., there is a large amount of overlap between the two sets), and indices near~0 indicating that few elements belong to both sets (i.e., there is little overlap between the two sets). 

Unlike the open magnetic flux and magnetic null points, the solar wind predictions are continuous, so the Jaccard index cannot be used. To quantify the agreement between predictions from a pair of boundary maps, we compute the fractional area in which the wind speed differs by less than 100\,km\,s$^-1$. This value is small enough to distinguish high speed from low speed wind, but large enough that fluctuations in particularly the slow wind will not be evaluated as differences between the models. Like the Jaccard index, this metric can take on values between 0 and 1 with values near 1 indicating the best agreement.

One focus in the comparisons is whether the difference in the photospheric open-flux regions, null points, sector boundaries, and wind speeds based on the ADAPT and SSW-PFSS models is greater than the differences in these various features in the group of realizations from the ADAPT model. We interpret the variations across the realizations of the ADAPT model as a way to characterize the uncertainties that result from the effect of convection on the surface magnetic flux in ADAPT. If the results of the two separate SFT models were to fall within the range of ADAPT realizations, we ascribe the differences as likely due to the random effects of convection; conversely when the differences between the two separate SFT models is larger, this can be ascribed to differences in the assumptions and implementation built into the two SFT modelling schemes.

\section{Results and Discussion}

We present here the results for each of the points of comparison in turn. 

\subsection{Sector Boundaries and Open Field Footpoints}\label{sec:openfield}

After illustrating the sector structure and open flux footpoints for one day, we show their variations over a solar cycle as well over an interval of a few days during which an active region is entering the assimilation window.

\begin{figure*}[ht!]
\includegraphics[width=\textwidth]{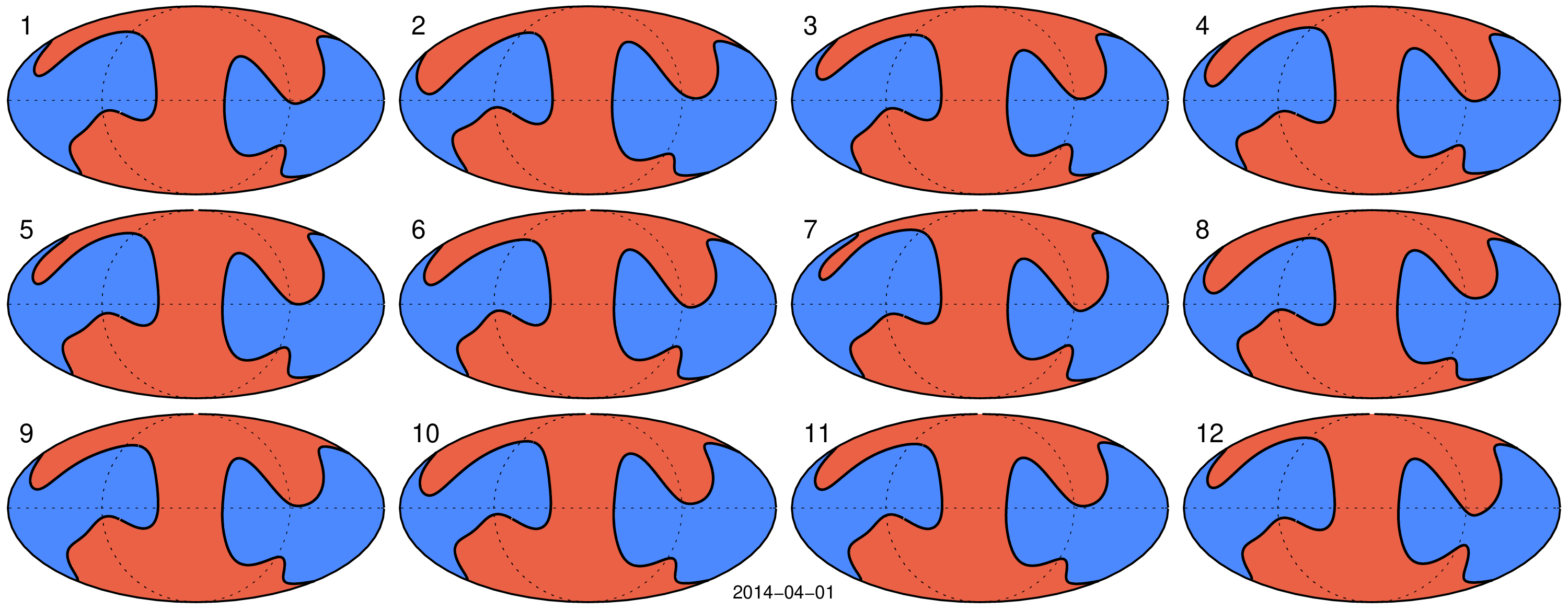}
\caption{Sector structure at a radius of 2.5~$R_\odot$ for the set of 12 ADAPT-HMI realizations for 2014 April~1, as determined by a PFSS extrapolation scheme. Sectors with positive polarities are red and sectors with negative polarities are blue, with the PILs that divide the sectors shown as a thick black line for each realization. As in Figure~\ref{fig:map-at-cbase}, Mollweide map projections are used, with the centroid longitude chosen to be the Carrington longitude as viewed from Earth for this date. In each map, the dashed circle indicates the solar limb as viewed from Earth, and the horizontal dashed line represents the solar equator. The sector structure is qualitatively similar for all of the ADAPT-HMI realizations. This figure is sampled from an animation (available online) consisting of one image per day for each day in April of even numbered years between 2012 and 2020. Around solar maximum, the animation shows a comparatively convoluted sector boundary, and sometimes more than one distinct boundary; near solar minimum, the sector boundary approximately follows the equator. At all times, there are only small variations among the different realizations.}
\label{fig:pil-at-ssurf-ADAPT-set}
\end{figure*}

\begin{figure}[ht!]
\includegraphics[width=\columnwidth]{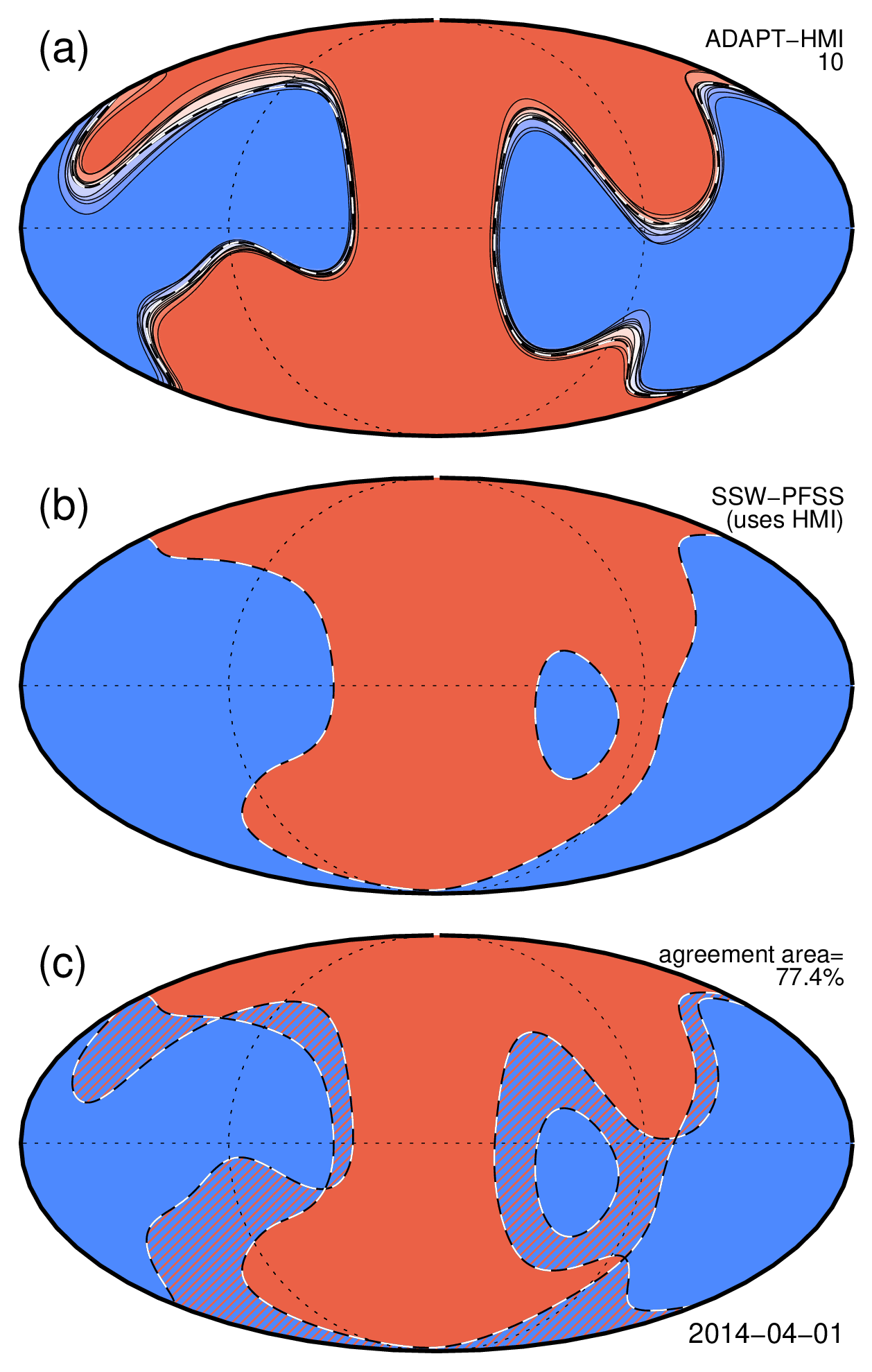}
\caption{Sector structure at a radius of 2.5~$R_\odot$ for 2014 April~1 for (a) the set of 12 ADAPT-HMI models and (b) the SSW-PFSS model. As in earlier figures, Mollweide projections are used. In panel~(a), the black-and-white dashed line represents the polarity inversion line corresponding to ADAPT map \#10, which is deemed to best represent the set of ADAPT realizations, as determined by examining the areas over which the sector polarities agree when compared to all other realizations in the set. In panel~(c), the polarity inversion lines from the representative ADAPT map and the SSW-PFSS map are both plotted. The solid red and blue areas indicate regions where the polarities of the sectors from both maps agree, and cover 77.4\% of the total area, whereas in the hatched regions the polarities are found to differ. In all three panels, sectors are shaded red or blue, depending on polarity (negative or positive, respectively).  This figure is sampled from an animation (available online) consisting of one image per day for each day in April of even numbered years between 2012 and 2020. Around solar maximum, the animation shows that the SSW-PFSS model predicts a comparatively convoluted sector boundary that can be morphologically quite distinct from the ADAPT-HMI prediction; near solar minimum, the sector boundary approximately follows the equator with relatively small variations among the different models.}
\label{fig:pil-at-ssurf-moll}
\end{figure}

\begin{figure}[ht!]
\includegraphics[width=\columnwidth]{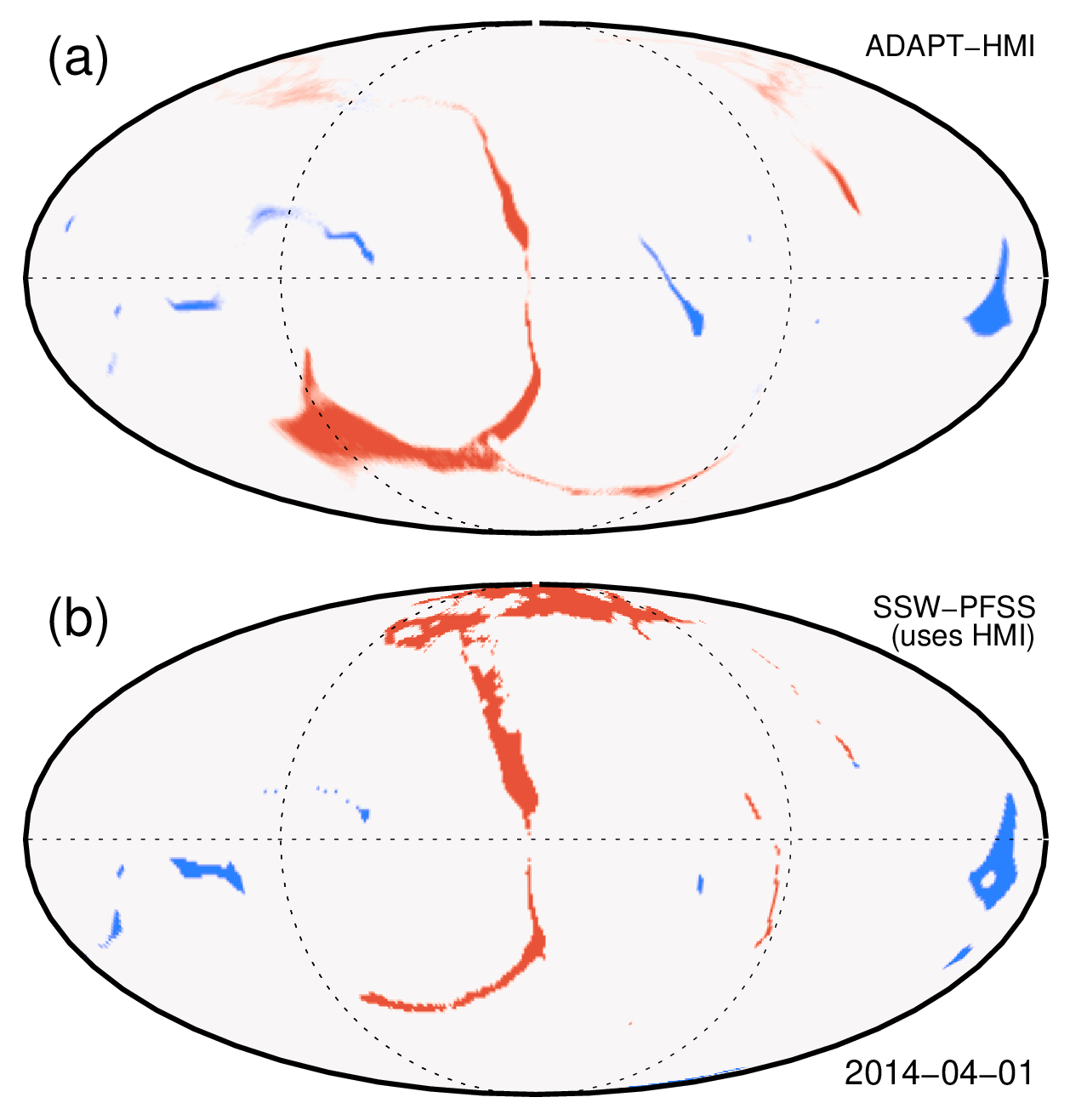}
\caption{Maps of open-flux regions at a radius of 1.02~$R_\odot$ as determined by applying potential-field extrapolations for 2014 April~1 to (a) the set of 12 ADAPT-HMI models and (b) the SSW-PFSS model. As in earlier figures, Mollweide projections are used. The open-flux regions are colored red or blue, depending on polarity (negative and positive, respectively). For the April 2014 time interval, while there is some similarity in the shapes of the open-flux regions, in many places (especially the polar regions) the regions differ significantly. We ascribe this effect to the weaker polar-cap fluxes found in the ADAPT-HMI model when compared with the SSW-PFSS model (cf., Fig.~\ref{fig:map-at-cbase}) during this time, resulting in a Jaccard index of 0.17 for this date. In other years, the Jaccard indices are more typically found to have values between 0.4 and 0.6 (approximately). This figure is sampled from an animation (available online) consisting of one image per day for each day in April of even numbered years between 2012 and 2020. Near solar maximum, the animation shows that PFSS models from both ADAPT-HMI and SSW-PFSS show trans-equatorial open-flux regions, typically in similar locations, although the exact location, size, and extent varies between the models; near solar minimum, the open-flux regions in both models are predominantly near the poles, although with slightly different extent.}
\label{fig:openflux-at-lowcorona}
\end{figure}

\begin{figure}[ht!]
\includegraphics[width=\columnwidth]{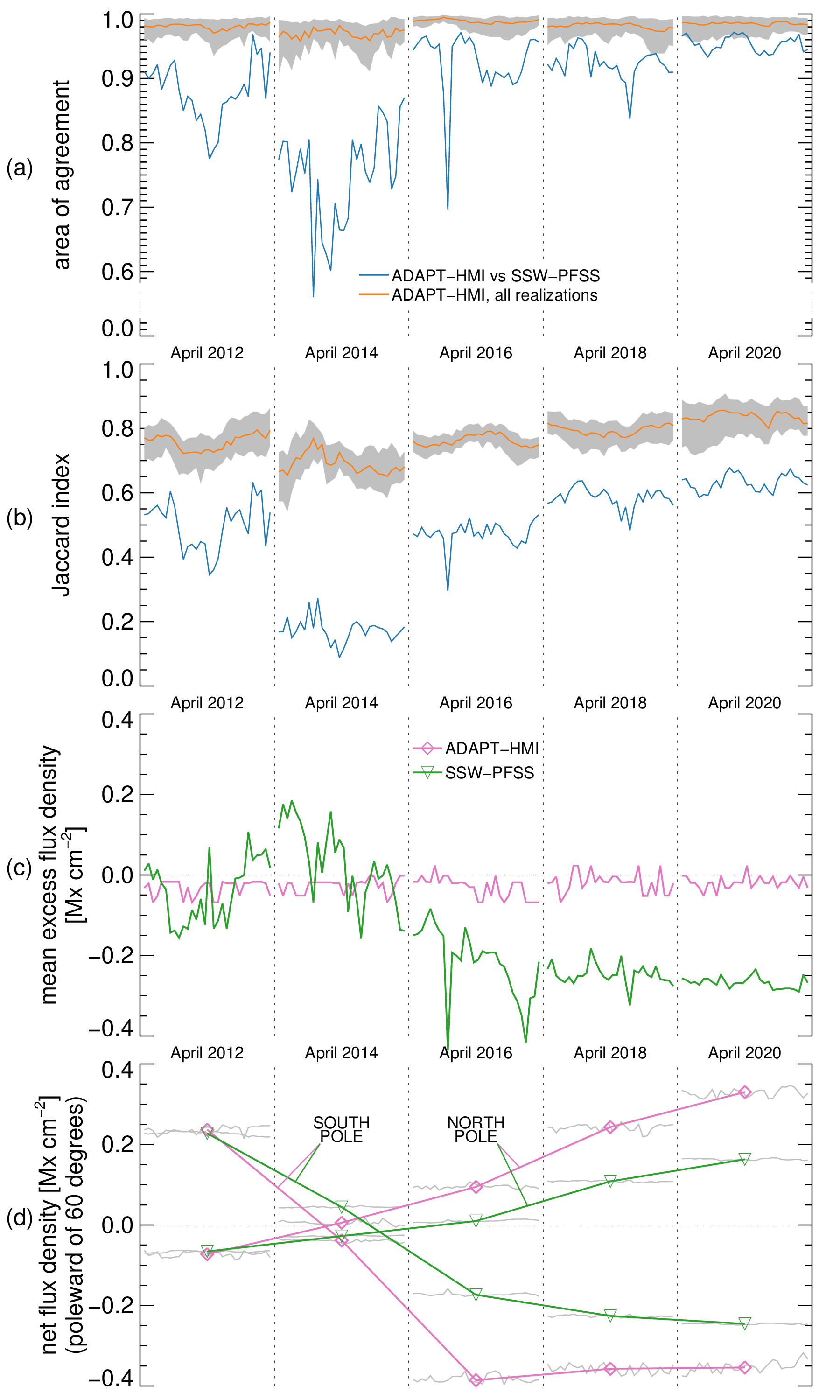}
\caption{Various quantities (one per day) for each date in April for all even-numbered years between 2012 and 2020, comprising (a) the areas of agreement when comparing the polarities in the sectors at 2.5\,R$_\odot$ from the representative ADAPT-HMI map with those from the SSW-PFSS map (blue line; cf., panel~(c) of Fig.~\ref{fig:pil-at-ssurf-moll}), or the median area of agreement when comparing of the polarities of the sectors at 2.5\,R$_\odot$ for each pair of realizations in the ADAPT-HMI set (orange line), (b) the Jaccard index when comparing the open-flux regions at a height near the base of the corona found from the representative ADAPT-HMI map with those from the SSW-PFSS map (blue line; cf., Fig.~\ref{fig:openflux-at-lowcorona}), or the median Jaccard index for each pair of realizations in the ADAPT-HMI set (orange line), (c) the mean excess flux density as integrated over the ADAPT-HMI and SSW-PFSS maps (magenta and green lines, respectively) that was removed before performing the comparison analysis, and (d) monthly averages of the net flux poleward of latitudes of $\pm$60$^\circ$ in each hemisphere for the ADAPT-HMI and SSW-PFSS maps (magenta and green lines, respectively). In panels (a) and (b), the shaded band indicates the min-max extent of the plotted quantity, and in panel (d) the gray points indicate the values used to calculate the monthly averages. The agreement among ADAPT-HMI realizations for both the sector structure and the open flux is consistently better than the agreement between ADAPT-HMI and SSW-PFSS.}
\label{fig:pil-at-ssurf-allmonths}
\end{figure}

\subsubsection{2014 April 1 as an Example}

We show in Figure~\ref{fig:pil-at-ssurf-ADAPT-set} the sector structure at a radius of 2.5~$R_\odot$ for the set of 12 ADAPT-HMI realizations for 2014 April~1. On this date, the polar fields are relatively weak when compared with other phases of the sunspot cycle, as the leftover polar-cap flux from the previous activity-minimum period has mostly been cancelled and is gradually being replaced by flux having the opposite polarity. The relatively weak axial dipole moment thus allows for both the equatorial dipole and higher-degree harmonics to influence the sector structure, which makes it more complex than at other phases of the sunspot cycle. Among the set of ADAPT-HMI realizations, the resulting sector-structure maps are qualitatively similar, especially in the Earth-facing hemisphere, however noticeable differences are evident toward the edges of the map projections at farside longitudes. 

We show in Figure~\ref{fig:pil-at-ssurf-moll} the sector structure at a radius of 2.5~$R_\odot$ for the ADAPT-HMI and SSW-PFSS models for 2014 April~1. In panel~(a) of the figure, all 12 maps in the ADAPT set from Figure~\ref{fig:pil-at-ssurf-ADAPT-set} are represented, with the shading at each location reflecting the degree to which the set of ADAPT maps agree on the polarities (darker colors indicate more agreement). The polarity inversion lines from each of the 12 realizations are drawn in the figure. Of the 12 ADAPT-HMI realizations, we define as representative of the set the one model with the least area of polarity mismatch, when compared with each of the others. For this date, this is map \#10, and the polarity inversion line from this map is represented in panel~(a) of the figure as a thicker, black-and-white dashed line. The representative ADAPT-HMI model changes from date to date.

In panel~(b) of Figure~\ref{fig:pil-at-ssurf-moll} the sector structure at 2.5~$R_\odot$ of the SSW-PFSS model is shown for the same date, with its polarity inversion line represented as a black-and-white dashed line. When comparing this map to the set of ADAPT-HMI maps, they appear qualitatively similar, however differences are evident at the south pole and at farside longitudes. In both of these locations, the photospheric magnetic field is less well constrained by the observations and is instead determined by the evolution of the photospheric fields incorporated into the flux-transport models. Because the polar-cap fields were in the process of changing sign for this date in 2014, we ascribe some of the differences in the sector structure for this date to  differences in the rate at which the models advect flux poleward. In 2014, these variations led to polar caps having different amounts of average flux, as is evident in Figure~\ref{fig:map-at-cbase}.

Panel~(c) of Figure~\ref{fig:pil-at-ssurf-moll} illustrates a comparison of the sector structures derived from ADAPT-HMI and SSW-PFSS. The 12 realizations from ADAPT-HMI are represented by the same map identified in panel~(a), i.e., the sector structure from this single map is used in the comparison. Over the full 4$\pi$ steradians, the polarities of the sectors agree in 77.4\% of this total area. The region for which the polarities are found to differ are confined to a band located between the two polarity inversion lines. This band is depicted as a blue-and-red hatched area in the figure, and covers the remaining 22.6\% of the total area. The figure also illustrates that the difference between ADAPT-HMI and SSW-PFSS is noticeably larger than the variation amongst the set of ADAPT-HMI maps (cf., Fig.~\ref{fig:pil-at-ssurf-ADAPT-set}), a result that will be shown in the next subsection to apply more generally to all of the dates considered in this study.

For a PFSS model, the magnetic field that intersects the source surface can be traced down through the model to the photosphere. Figure~\ref{fig:openflux-at-lowcorona} shows the regions at a radius representing the base of the corona through which the open field passes. The blue and red areas in each of the sectors shown in Figures~\ref{fig:pil-at-ssurf-moll}(a) and~(b) map down to the correspondingly colored regions shown in Figures~\ref{fig:openflux-at-lowcorona}(a) and~(b). As with the sector-boundary maps, the open-flux regions from ADAPT-HMI and SSW-PFSS appear qualitatively similar, with both models possessing a mostly north-south aligned narrow band of open field generally aligned with the central meridian, and a patch of opposite-polarity open flux near the right (eastern) edge of the figure. Many of the smaller open-flux regions have a correspondence between the two models. Differences are evident in the exact boundaries of the open-flux areas that are present in both models, and there are some open-flux regions that are present in only one of the types of models. There is a particularly pronounced open-flux region near the north pole in the SSW-PFSS model that is absent in ADAPT-HMI. 

To quantify the degree to which the open-flux maps from the representative ADAPT-HMI model and the SSW-PFSS model overlap, we calculate their Jaccard similarity index, defined in \S\ref{sec:points-comparison}. Though the polarities of the open-flux regions are depicted in Figure~\ref{fig:openflux-at-lowcorona} as either red and blue, the sign of the polarities is ignored when calculating Jaccard indices, because we do not find any locations for any date in the study for which the magnetic field is open in both the representative ADAPT-HMI model and the SSW-PFSS model but the polarities differ.

For the 2014 April~1 example, the Jaccard similarity index is 0.17, which indicates that there is not a lot of overlap in the open-flux regions. We mainly ascribe this low value to differences in the polar-field strength between the ADAPT-HMI and SSW-PFSS models, since this property strongly affects the global field connectivity, and for example results in a large area of open flux at the north pole in the SSW-PFSS model that is not present in the ADAPT-HMI model. However, we also note that this index is sensitive to the exact location of open field, so even when open flux is present in both models but at slightly different locations, the Jaccard index will be small. During other phases of the Sunspot Cycle~24, Jaccard indices are higher and therefore indicate a greater degree of overlap,  typically falling between~0.4 and~0.5, as discussed in the following subsection.

\subsubsection{Trends Throughout Sunspot Cycle 24}

To gain intuition into how the ADAPT-HMI and SSW-PFSS maps compare throughout a sunspot cycle, we plot in Figure~\ref{fig:pil-at-ssurf-allmonths} several quantities of interest, generated from daily maps from the ADAPT-HMI and SSW-PFSS models, over the months of April in even-numbered years from 2012 to 2020. In this manner, we capture variations on timescales of both a few days and over a solar cycle.

Panel~(a) of Figure~\ref{fig:pil-at-ssurf-allmonths} shows as a blue line the area of agreement in the sector structure at 2.5~$R_\odot$, as calculated using PFSS extrapolations from the representative ADAPT-HMI map and from the SSW-PFSS map. These areas of agreement are plotted as a fraction of the total area of 4$\pi$ steradians. The range of agreement areas across the set of ADAPT-HMI maps is plotted as a gray band, with the median agreement area in orange. These agreement areas for the set of ADAPT-HMI maps are typically very close to unity and lead to the result that there is not a large difference in the sector-boundary maps determined from the set of 12 ADAPT-HMI realizations,whereas there is a much smaller area of agreement between ADAPT-HMI and SSW-PFSS.

The blue line in panel~(b) plots the Jaccard indices of the open-flux regions as calculated using PFSS extrapolations from daily maps from the ADAPT-HMI and SSW-PFSS models. We compare these open-flux regions at a height that nominally represents the base of the corona. Also shown in orange are the median Jaccard indices for the open-flux regions as determined from each pair of ADAPT-HMI extrapolations. As with the sector-boundary agreement area metric, we find that the correspondence between the various ADAPT-HMI models is greater than that found when comparing the ADAPT-HMI models to the SSW-PFSS models. The gray band demarcates the range of Jaccard indices for the set of ADAPT-HMI maps.

Even considering the differences between ADAPT-HMI and SSW-PFSS, throughout much of Sunspot Cycle~24, there is between approximately 80\% and 90\% agreement in the polarities of the sectors at 2.5~$R_\odot$, and the Jaccard index for the open flux is found to fall mainly in the 0.4--0.6 range. The exception is the month of April 2014, during which the agreement is evidently lower. The best agreement occurs in April 2020. 

To understand this trend, consider panels~(c) and~(d) in Figure~\ref{fig:pil-at-ssurf-allmonths}, in which are respectively shown for both ADAPT-HMI and SSW-PFSS the amount of monopole flux removed via the scheme described in Section~\ref{sec:monopole} and the integrated polar-cap flux poleward of latitudes of $\pm$60$^\circ$ in each hemisphere. As we have noted, the polar fields are in the process of reversing during 2014, and yet the reversals in ADAPT-HMI and SSW-PFSS are shifted in time, such that in 2014 the south polar cap field is still positive for SSW-PFSS while it has already switched to negative for ADAPT-HMI. The converse is the case for the northern polar cap, although the difference in polar-cap flux is smaller in both models. Even though the fields are not so different in magnitude at this time compared with later years analyzed, this difference in polarity has a substantial impact on the large-scale geometry of the coronal magnetic field, which affects the sector-structure and the open-flux footpoint metrics used as points of comparison in this study.

\begin{figure*}
    \centering
    \includegraphics[width=\textwidth]{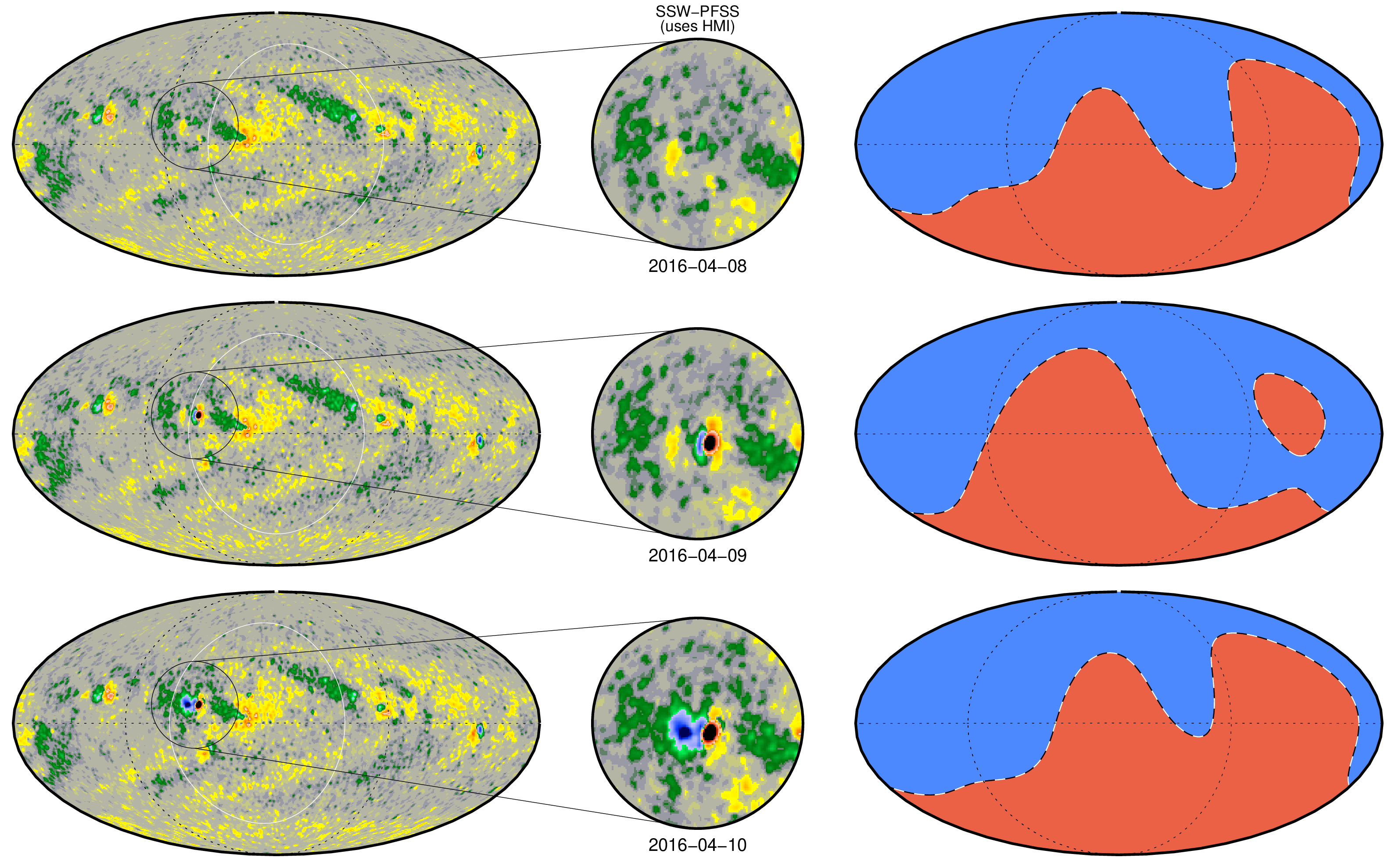}
    \caption{A sequence of SSW-PFSS maps from April 2016 showing the effects of the partial assimilation of an active region. The left-most column shows the flux density at a radius of 1.005~R$_\odot$ (as in Fig.~\ref{fig:map-at-cbase}) on three consecutive days (2016 April~8--10). The white oval in each map shows the approximate boundary of the assimilation window for the SSW-PFSS map, which is sweeping eastward as each day passes. In the black circle, an active region (AR~12529) appears, first partially on 2016 April~9 as only the leading polarity falls within the assimilation window, and then finally as a completely assimilated region on the next day. The right-most column shows the corresponding sector structure at a radius of 2.5~R$_\odot$ (as in Fig.~\ref{fig:pil-at-ssurf-moll}). On 2016 April~9, there is a drastic, but transient, change in the sector structure due to the anomalously large flux imbalance on that day caused from the partially assimilated active region.}
    \label{fig:assimilation-transient}
\end{figure*}

In addition to the solar cycle trend, there are transient intervals when the agreement between ADAPT-HMI and SSW-PFSS is temporarily worse. Often, these more temporary intervals 
correspond to instances where only part of an active region falls within the assimilation window for the SSW-PFSS model. One indicator of this effect is the flux associated with the monopole, a quantity that is removed from each model prior to the comparison analysis. For example, during the 2016 April~8--10 interval, the PFSS-SSW monopole temporarily spikes  (cf., panel~(c) of Fig.~\ref{fig:pil-at-ssurf-allmonths}) due to the partial assimilation of an active region. This situation is illustrated in  Figure~\ref{fig:assimilation-transient}, in which maps of the near-photospheric flux and the sector-boundary structure are shown for those three consecutive days from the SSW-PFSS model. In this particular case, comparing the sector-boundary structure on April ~8 and~10 indicates that the addition of the fully assimilated active region to the PFSS-SSW model did not greatly perturb the global structure of the field, due to its orientation being aligned with the large-scale pre-existing field. Therefore, we conclude that the transient changes seen in the metrics plotted in Figure~\ref{fig:pil-at-ssurf-allmonths} for 2016 April~9 are due primarily to the unbalanced flux associated with the half-assimilated region on that date. Because ADAPT-HMI uses a smoothly varying weighting function based on distance from disk center in determining the degree to which new data are assimilated (longitudes farther away from the central meridian are weighted less), abrupt changes in the monopole correction occur less often in the ADAPT-HMI model.  

\subsection{Wind Speed Predictions}\label{sec:windspeed}

\begin{figure}[ht!]
\includegraphics[width=\columnwidth]{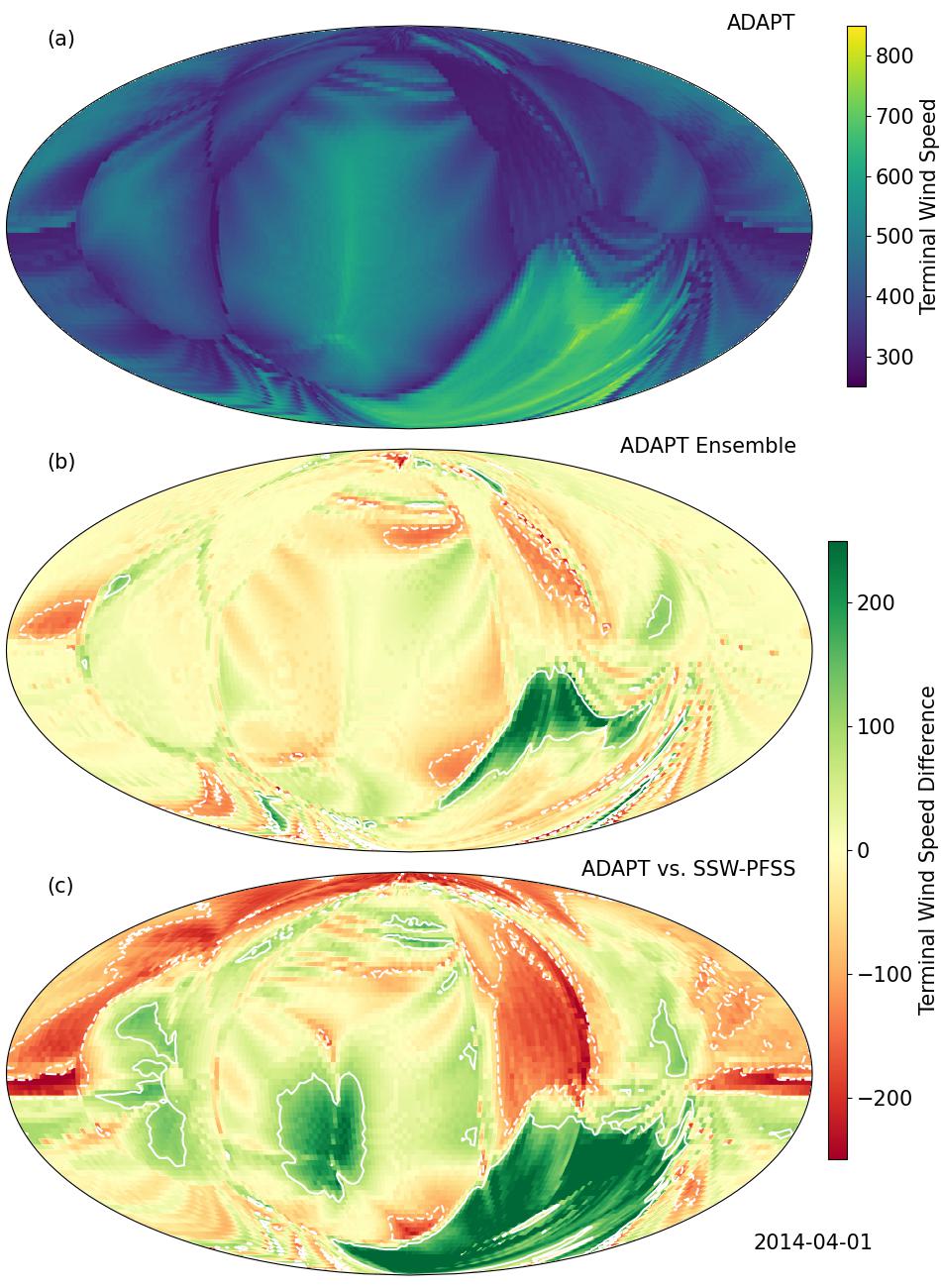}
\caption{Solar wind speed predictions from the WSA model at 5.0\,$R_\odot$ for 2014 April~1, shown in a Mollweide projection. Panel (a) shows the wind speed predicted by a representative ADAPT-HMI realization. Panel (b) shows the median difference between the wind speed predicted from the representative ADAPT-HMI realization and all the other realizations. Panel (c) shows the difference between the wind speed predicted from the representative ADAPT-HMI realization and the SSW-PFSS model. In panels (b) and (c), the contour encloses differences of at least 100\,km\,s${}^{-1}$, the threshold used to determine the area in which there is reasonable agreement between wind speed predictions. The extent of the high speed wind area [light green in panel (a)] generated from the ADAPT-HMI boundary condition varies between realizations [much smaller dark green area in panel (b)], but is entirely absent in the predictions from the SSW-PFSS model (large area of dark green in panel (c)], due to differences in the polar field. This figure is sampled from an animation (available online) consisting of one image per day for each day in April of even numbered years between 2012 and 2020. Around solar maximum, the animation shows a mix of areas of higher and lower wind speed predictions, across a range of latitudes, from the SSW-PFSS boundary condition compared with ADAPT-HMI; around solar minimum, wind speed predictions from the ADAPT-HMI boundary condition tend to be higher than from SSW-PFSS, particularly at lower latitudes.}
\label{fig:comp_wind_speed}
\end{figure}

Figure~\ref{fig:comp_wind_speed} illustrates the differences in the wind speed predicted by the WSA model for 2014 April~1 when driven from boundary conditions produced by ADAPT-HMI versus SSW-PFSS. Panel (a) shows the predicted wind speed at 5.0\,$R_\odot$ from a representative ADAPT-HMI map. On this date, there is predominantly slow wind present, with the main region of fast wind being in the southern hemisphere. The differences between the predictions from different realizations of ADAPT-HMI [panel (b)] are generally less than 100\,km\,s${}^{-1}$, with the largest differences occurring close to the boundaries between low and high speed wind. Comparing the predictions from a representative ADAPT-HMI map with those from SSW-PFSS [panel (c)], the differences in the majority of the area are still less than 100\,km\,s${}^{-1}$, but are more widespread than between ADAPT-HMI realizations. In particular, the entire area of fast wind in the southern hemisphere has large differences, indicating that the SSW-PFSS model does not produce significant high speed wind in the southern hemisphere, while in the northern hemisphere, the SSW-PFSS model produces areas of high speed wind that have no counterpart in the ADAPT-HMI realizations. As noted previously, on this date there are substantial differences in the polarity of the polar fields, resulting in, for example, a large area of open flux at the north pole in the SSW-PFSS model that is not present in any of the ADAPT-HMI realizations (c.f., Fig.~\ref{fig:openflux-at-lowcorona}). This open flux, present only in the SSW-PFSS model, is the source of the high speed wind not seen in the ADAPT-HMI model.

\begin{figure}[ht!]
\includegraphics[width=\columnwidth]{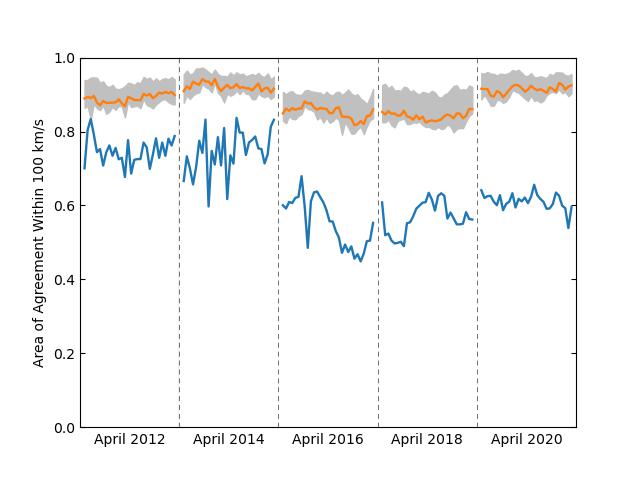}
\caption{The fractional area at 5.0\,$R_\odot$ where the absolute value of the difference in the wind speed predicted from two different boundary maps is less than 100\,km\,s${}^{-1}$ for all dates in April of even numbered years between 2012 and 2020. The results from a representative ADAPT-HMI map are compared with the result from the SSW-PFSS model (blue) as well as the median fractional area from the ADAPT-HMI maps (orange); the shaded band indicates the extent of the variations among the ADAPT-HMI realizations. Throughout, there is much more consistency among the wind speed predictions from the ADAPT-HMI ensemble members than between the predictions from SSW-PFSS and ADAPT-HMI, although the greatest differences are seen in 2016, 2018, and 2020.} 
\label{fig:solar_cycle_wind}
\end{figure}

This spatial pattern of the differences between the predictions from ADAPT-HMI realizations being concentrated near the boundaries between fast and slow wind holds throughout the solar cycle, as does the more widespread disagreement between the predictions from the SSW-PFSS model and ADAPT-HMI, however the amplitude varies. To quantify this, we show in Figure~\ref{fig:solar_cycle_wind} the fraction of the area for which the wind speed predictions from two different boundary conditions differ by less than 100\,km\,s${}^{-1}$. The predictions from the ADAPT-HMI realizations typically have differences of less than 100\,km\,s${}^{-1}$ over 90\% of the area, and over at least 80\% of the area for all the dates we evaluated. The differences in the predictions from SSW-PFSS compared with ADAPT-HMI show much greater variability, ranging from good agreement over more than 80\% of the area to less than 50\% of the area. In 2012 and 2014, the agreement between the ADAPT-HMI and SSW-PFSS wind predictions is much better than in 2016, 2018, and 2020 dropping from around 75\% of the area having wind speeds within 100\,km\,s${}^{-1}$ to around 55\% of the area having wind speeds within 100\,km\,s${}^{-1}$. In addition to these long term variations, like the sector structure, there is evidence for large (10\% of the area), transient differences in the wind speed predictions at times when an active region is entering the SSW-PFSS assimilation window.

The SSW-PFSS polar field strengths for 2016, 2018, and 2020 are smaller than the polar field strengths for ADAPT-HMI c.f., Fig.~\ref{fig:pil-at-ssurf-allmonths}(d)]. This results in more low-latitude open field regions in SSW-PFSS, particularly for the northern hemisphere, and particularly in 2016. Thus we posit that the solar cycle dependence of the wind speed differences is due to the stronger polar fields in ADAPT-HMI in the later years resulting in overall more high speed wind because the wind is coming preferentially from the polar coronal holes as compared with SSW-PFSS.

\subsection{Coronal Magnetic Null Points}\label{sec:nullpoints}

Magnetic null points were located in each model down to a radius of $1.016\,R_\odot$, corresponding to one grid point above the lower boundary, and the clustering algorithm (\S\ref{sec:nullmatch}) was applied to all these null points. However, in imposing a minimum radius for the null points, there is a risk that a null point in one model belonging to a cluster may lie just above the threshold, while a null point in another model that should belong to the same cluster may lie just below the threshold and is thus not associated with the cluster. To reduce the impact of this effect, results are generally only shown for clusters with a minimum radius of $1.05\,R_\odot$. Individual null points may have a smaller radius than this, provided they are assigned to a cluster with a radius greater than the threshold.

\begin{figure}[ht!]
\plotone{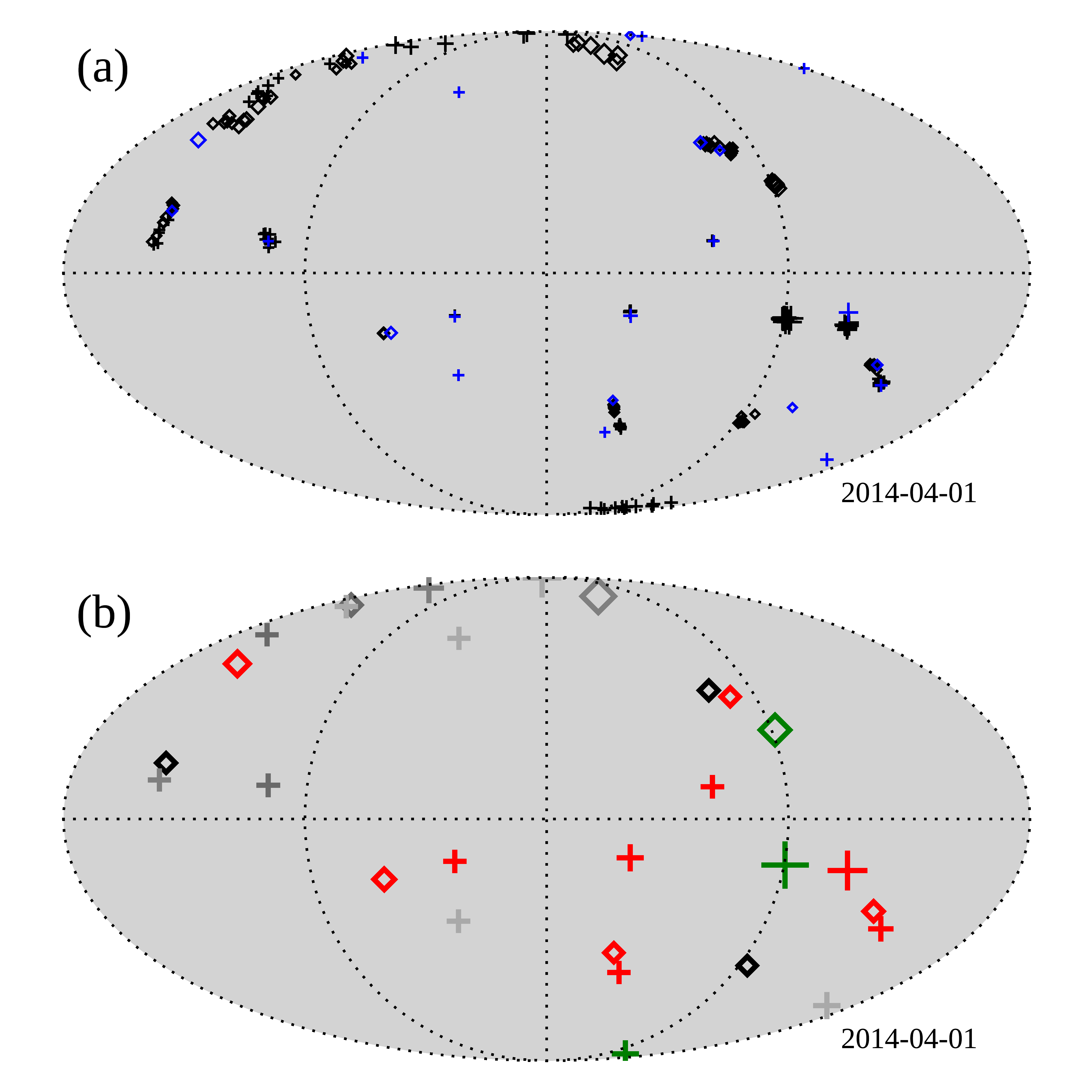}
\caption{Null points on 2014 April~1, shown in a Mollweide projection. Panel (a) shows all the null points found for the different models. Black symbols indicate null points that are present in the PFSS extrapolation for any realization of the ADAPT-HMI model; blue symbols indicate null points in the SSW-PFSS model. Positive and negative nulls are shown with pluses and diamonds respectively, while the symbol size is proportional to the radius of the null (i.e., large symbols indicate nulls that are at a large radius). Only null points assigned to a cluster with radius larger than $1.05 R_\odot$ are shown for clarity. Some individual nulls may lie below this radius, provided the radius of the mean null location for the cluster is above the threshold. Panel (b) shows the corresponding position of each cluster mean, with the symbol color determined by the number of (non-ghost, see Appendix) nulls in the cluster. Red indicates a null that is present in the PFSS extrapolation for all 12 realizations of ADAPT-HMI plus SSW-PFSS, green indicates a null that is present in the PFSS extrapolation for all 12 realizations of ADAPT-HMI, but \textit{not} SSW-PFSS, while shades of black to grey indicate decreasing numbers of nulls in the cluster. This figure is sampled from an animation (available online) consisting of one image per day for each day in April of even numbered years between 2012 and 2020. Around solar maximum, the animation shows null points present in the PFSS model from all the ADAPT-HMI realizations more frequently than null points are present in the PFSS model from one realization of ADAPT-HMI and SSW-PFSS across all longitudes; near solar minimum, the null points in the SSW-PFSS model are largely indistinguishable from the null points in the PFSS model from a realization of ADAPT-HMI, and there is noticeably more agreement among the models in the presence of null points above areas where data have recently been assimilated into the models.}
\label{fig:nulls}
\end{figure}

Figure~\ref{fig:nulls} shows an example of the results of the null points found on 2014 April~1. About half of the null points are present in the PFSS model from the SSW-PFSS model and all the realizations of ADAPT-HMI. However, all possible exceptions to this occur. Slightly below and left of the center of the image is a positive null point found only in the SSW-PFSS model (blue symbol in the top panel; light grey symbol in the bottom panel). Slightly north of the equator and slightly more than half way to the left of the center of the figure is a positive null point present in the SSW-PFSS model and most but not all of the realizations of ADAPT-HMI (mix of blue and black symbols in the top panel; grey symbol in the bottom panel). Further to the left of the figure is a positive null point present in most but not all of the realizations of ADAPT-HMI, but not present in the SSW-PFSS model (only black plus symbols are present in the top panel). Near the south pole (bottom of the image) is a null point present in the PFSS model from all of the ADAPT-HMI realizations but not the SSW-PFSS model (green symbol in the bottom panel).

\begin{figure}[ht!]
\includegraphics[width=\columnwidth]{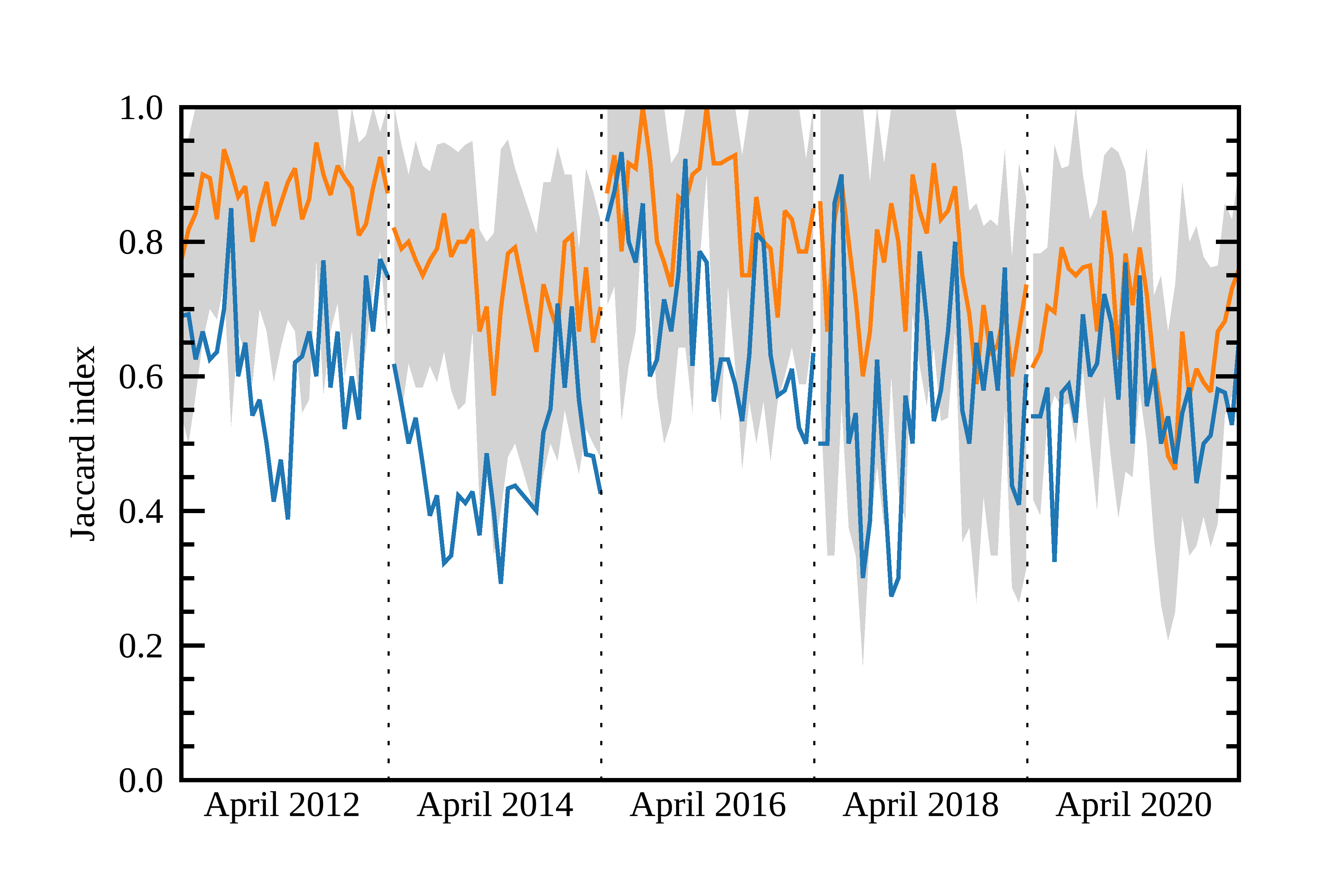}
\caption{The Jaccard index for the null points present in the PFSS extrapolation from a representative realization of the ADAPT-HMI model and the SSW-PFSS model (blue), and the median Jaccard index for null points in the PFSS model from the representative ADAPT-HMI realization and all the other realizations (orange) for all dates in April of even numbered years between 2012 and 2020; the shaded band indicates the extent of the variations among the ADAPT-HMI realizations. The reference realization is the one with the maximum agreement with the other realizations. Only null points whose average radius is greater than 1.05\,$R_\odot$ are included. Null points are much more likely to be present in both the PFSS extrapolations from a pair of ADAPT-HMI maps than in both the PFSS extrapolation from the representative ADAPT-HMI map and the SSW-PFSS model approaching solar maximum, but by solar minimum, the SSW-PFSS nulls points cannot be clearly distinguished from the null points in the PFSS model from an ADAPT-HMI realization.
}
\label{fig:solar_cycle_nulls}
\end{figure}

To quantify the overall agreement between the null points in the PFSS extrapolation from the different SFT models, the Jaccard index for different pairs of models is shown in Figure~\ref{fig:solar_cycle_nulls}. A minimum radius of $r=1.05\,R_\odot$ on the average radius of the null is again imposed to avoid including large numbers of low-lying nulls that can be sensitive to noise in the boundary maps, and are not likely to be physically meaningful. On average, the PFSS models from any two realizations of the ADAPT-HMI model have approximately 70\% of the null points in common.  For the SSW-PFSS model, this drops to approximately 60\% of the nulls also present in the PFSS model from a representative ADAPT-HMI model.

There are solar cycle variations to the agreement among null points, with the best agreement occurring during 2012 and 2016. In 2020, the SSW-PFSS model behaves essentially like another realization of the ADAPT-HMI model, with virtually the same fraction of nulls present in both. We hypothesize there are two effects that explain this behavior: The polar field differences near solar maximum, and the increased importance of small spatial scale fields near solar minimum. Examining the animation of the null points, one sees that during 2020, the majority of null points above the Earth-facing side of the Sun are present in all the models, whereas null points above the far hemisphere tend to only be present in a few of the models. This is consistent with the null points being largely determined by relatively small spatial scale features, which are virtually the same between models within the assimilation window, but evolve in different ways outside the assimilation window.

Unlike the other measures, there is no clear evidence for transient variations in the null points while an active region is entering the assimilation window. It is likely when there is an individual null point associated with the active region being assimilated, it will not be present in both the SSW-PFSS and the ADAPT models, but the existence of other null points is largely unaffected. Note that this is partly a consequence of the way the monopole term is removed, which preserves the locations of polarity inversion lines, and should not be taken as a general property of different SFT models.

\section{Concluding Remarks}

Our results show that the choice of SFT model can greatly impact the results of global coronal and heliospheric models, based on the manner in which the different SFT models estimate the polar fields and how they assimilate new active regions. Large but transient differences are seen in most of the model predictions when an active region is crossing the edge of the assimilation window, specifically, during the time when there is substantial magnetic flux that has been assimilated by one model but not by another because of the different locations of the assimilation window. The impact of an active region being assimilated is not always localized to the corona in the vicinity of the active region but, with the exception in this study of the null points, can have global effects. 

There is also a solar cycle dependence that is different for each of the metrics considered, but in all cases, for the majority of the cycle, there is a greater difference between the predictions from a representative ADAPT-HMI model and the SSW-PFSS model than the differences among the predictions from different realizations of ADAPT-HMI. Thus we conclude that the assumptions about and the implementation of the effects of both large and small-scale flows in the SFT model are more impactful than the choice of a particular realization of the supergranulation in areas of the Sun that are not routinely observed.

The only way to completely avoid these impacts is to have data from different viewpoints of the Sun be assimilated into the SFT models. For the slowly evolving polar fields, it may be that polar observations separated by intervals of many months or even years would be sufficient to constrain the SFT models. This could be accomplished by (infrequent) assimilation of the polar observations or utilizing coronal hole observations to constrain the modeled polar flux distribution \citep[see][]{Schonfeld2022}, although additional tuning of the meridional flow profile and the treatment of supergranules to match the oberved polar flux might be sufficient. For the active regions, continual monitoring of all solar longitudes is needed, ideally in the form of magnetic field observations, but the use of helioseismology \citep{arge13} could be used to improve the results. In the context of synoptic maps, the impact of including observations from multiple Lagrange points has been modeled by \cite{Petrieetal2018,Pevtsovetal2020}, and shown to produce substantial improvements.

For scientific studies that rely upon SFT models for the input boundary conditions, we recommend avoiding intervals during which a new active region is rotating into the visible hemisphere from the far side of the Sun. Depending on the quantity of interest, some intervals of the solar cycle may be preferable. For example, times near the reversal of the polar fields lead to highly variable predictions for the sector structure. For more operational studies, when a prediction must be produced at all times, it is important to recognize that the level of uncertainty is much higher than is captured in the variations due to the treatment of supergranulation in a single SFT model. Ensembles of models should include representatives of multiple SFT models to capture the uncertainty in the polar fields.

\ \\

\noindent
The material presented here is based upon work supported by NASA grant 80NSSC19K0087 to NorthWest Research Associates, Inc. The color table used in Figures~\ref{fig:pil-at-ssurf-moll} and \ref{fig:openflux-at-lowcorona} is based on the CET\_D9 diverging color table available from the open-source Colorcet library, which is part of the HoloViz collaboration (\url{https://holoviz.org}). The ADAPT model development is supported by Air Force Research Laboratory (AFRL), along with AFOSR (Air Force Office of Scientific Research) tasks 18RVCOR126 and 22RVCOR012. This work utilizes data produced collaboratively between AFRL and the National Solar Observatory (NSO). The views expressed are those of the authors and do not reflect the official guidance or position of the United States Government, the Department of Defense (DoD) or of the United States Air Force. The appearance of external hyperlinks does not constitute endorsement by the DoD of the linked websites, or the information, products, or services contained therein. The DoD does not exercise any editorial, security, or other control over the information you may find at these locations.
We thank the referee for helpful comments that greatly improved the presentation of our results.

\appendix

\section{Differential Rotation and Meridional Flow Profiles} \label{sec:flow-profiles}

The differential rotation profile used in both the SSW-PFSS and ADAPT-HMI SFT models is taken from Table 1 of \citet{komm1993a} for the one-dimensional cross-correlation analysis applied to Kitt Peak magnetogram data from 1975--1991. The sidereal rotation rate $\Omega$ as a function of latitude $\lambda$ is
\begin{equation}\label{eqn:diff_rot}
    \Omega(\lambda) = A + B \sin^2(\lambda) + C \sin^4(\lambda),
\end{equation}
where the coefficients are
\begin{equation}\label{eqn:diff_rot_coef}\begin{aligned}
A&=\text{14$\fdg$42~day$^{-1}$}\\
B&=\text{--2$\fdg$00~day$^{-1}$}\\
C&=\text{--2$\fdg$09~day$^{-1}$}.
\end{aligned}\end{equation}
Because both the SSW-PFSS and ADAPT-HMI SFT models use a Carrington frame of reference, a solid-body sidereal Carrington rotation rate of 14$\fdg$18~day$^{-1}$ is subtracted off from $A$, giving $A$ a value of 0$\fdg$24~day$^{-1}$.
 
\begin{figure}[ht!]
\plotone{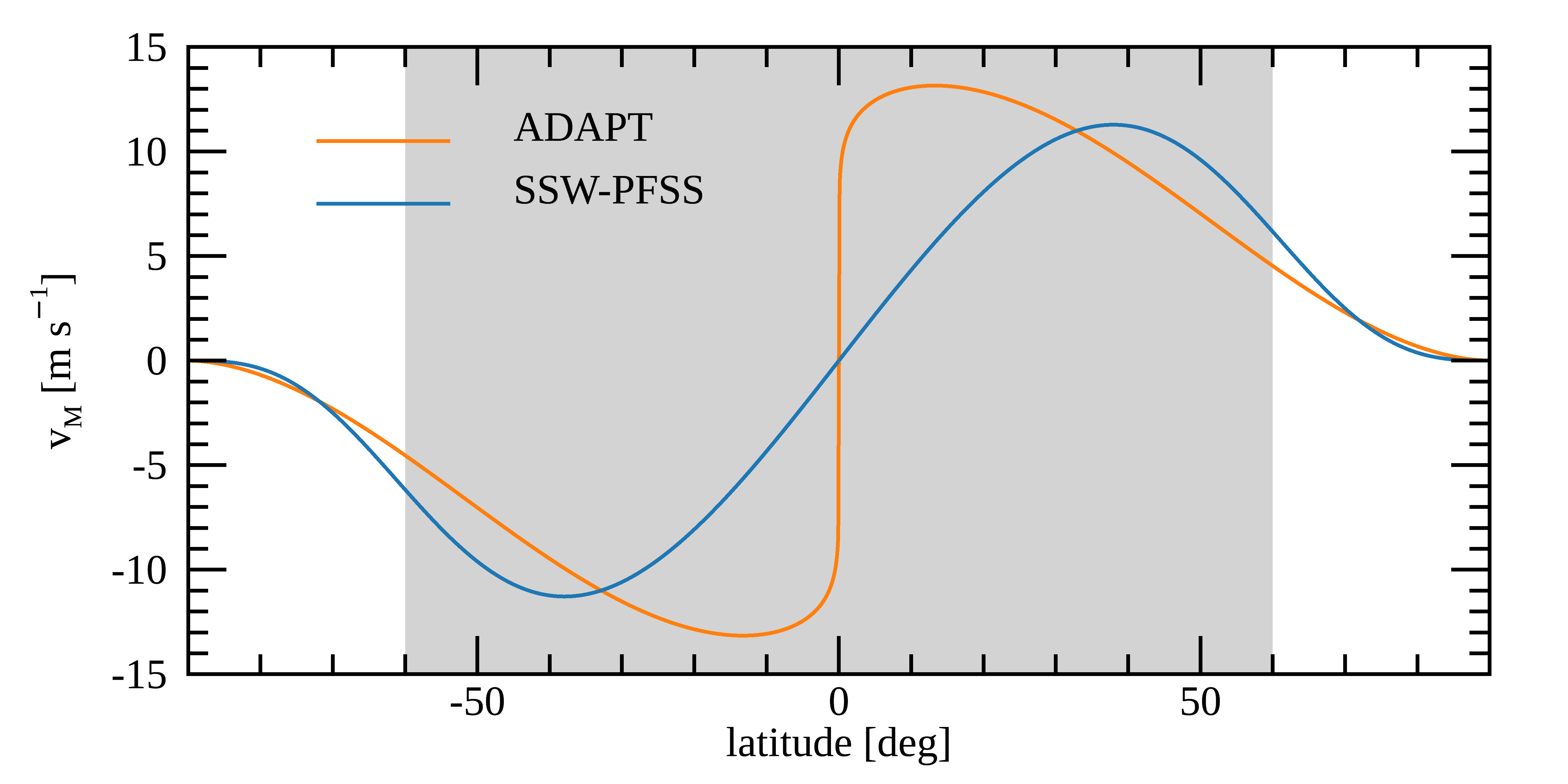}
\caption{Meridional flow profiles for the SSW-PFSS and ADAPT SFT models. Differences at low latitudes do not greatly impact the models because the data are assimilated in those regions, but differences at high latitudes likely contribute to the differences in the polar fields. The shaded area indicates the approximate latitude range in which new data are assimilated.}
\label{fig:meridional}
\end{figure}

The meridional flow profile $M(\lambda)$ used in the SSW-PFSS SFT model is based on a traditional $M(\lambda) \propto \sin(2\lambda)$ fit as measured in, e.g., \citet{komm1993b}, but with a tapering function applied to the polar latitudes. In terms of the colatitude $\theta$, the meridional flow profile is implemented as
\begin{equation}
    M(\theta)= v_A \sin(2\theta) f(\theta) f(\pi-\theta) 
\end{equation}
where the tapering functions $f(\theta)$ and $f(\pi-\theta)$ only apply poleward of 40$^\circ$ of latitude in each hemisphere. This function $f$ is
\begin{equation}
    f(\theta)=1-e^{-a \theta^3}.
\end{equation}
The constants $v_A$ and $a$ have values of 12.7\,m\,s$^{-1}$ and 3.0, respectively. The flow is poleward in both the northern and southern hemispheres, and peaks at 11.1\,m\,s$^{-1}$ at latitudes of about $\pm$35$^\circ$. When the tapering function is not included, the modeled polar cap flux becomes overly concentrated at the poles, and does not extend away from the poles as far as is typically observed during solar minima. We note that this $M$ profile is functionally similar to that given in Equation~(3) of \citet{schr2001b} but with different values of $v_A$ and $a$.

ADAPT-HMI uses a different meridional flow profile that follows the form given by \cite{WangRobbrechtSheeley2009}, namely 
\begin{equation}
M(\theta) = (16\, \textrm{m\,s}^{-1}) \vert \sin(\theta) \vert^{0.1} \vert \cos(\theta) \vert^{1.8}.
\end{equation}
The flow has a higher peak value of 13.2\,m\,s$^{-1}$ at a much lower latitude of about $\pm13^\circ$ when compared with the profile used by the SSW-PFSS model. Although this profile is not consistent with observations at lower latitudes, the regular assimilation of observational data into ADAPT-HMI means that only the high latitudes $(|\theta| \ga 65^\circ)$ are significantly affected by the meridional flow.

Figure~\ref{fig:meridional} shows the meridional flow profiles for the two models. The impact of the meridional flow on the polar flux is mainly driven by the slope of the flow profile between 60 and 70 degrees, where the poleward assimilation boundary resides depending on the observational orbital position. Everything model driven below 60 degrees, indicated by the shaded area in the figure, is dominated by supergranular flows. That is, new observations are assimilated at least once per rotation, a timescale on which the meridional flow has very little impact compared with the supergranulation in the region with the largest difference between the meridional flow profiles. The smaller differences at high latitudes are likely to contribute to the differences in the polar fields between the two models.

\section{Matching Null Points}\label{sec:nullmatch}

To associate null points in the PFSS model for different boundary conditions, a clustering algorithm based on $k$-means \citep{FeigelsonBabu2012} was used, where each cluster is considered to be a single null point present in one or more models. The approach of $k$-means clustering is to iteratively reassign each object (null point in this case) to a cluster such as to minimize the square of the within cluster distance
\begin{equation}
W = \sum_{j=1}^k \sum_{i=1}^{n_j} (\mathbf{x}_{ji} - \overline{\mathbf{x}}_j)^2
\end{equation}
where $\mathbf{x}_{ji}$ is the position of object $i$ in cluster $j$, $\overline{\mathbf{x}}_j$ is the mean position of the objects in cluster $j$, and $n_j$ is the number of objects in cluster $j$. In its iterative form, $k$-means clustering can easily fail to find the global minimum of $W$, so a simulated annealing algorithm \citep[e.g.,][]{Kirkpatrick1983} was implemented to search for the minimum. 

In order to prevent two null points present in the same model from being assigned to the same cluster, ``ghost'' null points are introduced to ensure that each model has the same number of null points. By construction, the ghost null points do not contribute to the within cluster distance, but they can be exchanged between clusters. Each cluster contains exactly one null point from each model although some of these may be ghost nulls. The simulated annealing algorithm exchanges null points from the same model between two clusters, evaluates the change in $W$, and decides whether to accept the exchange based on the Metropolis condition.

The additional challenge with $k$-means clustering is that the number of clusters must be specified a priori. A lower bound on this is the maximum number of null points of a given sign in a model. The algorithm is initiated with this number of clusters and allowed to converge. The $z$-score for the distance for each null point to its cluster center is then computed, and the number of clusters is increased by the number of $z$-scores greater than a threshold of 1. This is repeated until no cluster has a $z$-score greater than the threshold. This approach prevents clusters from containing a single null point that is much farther from the cluster center than any of the members of the cluster. A git repository of the null point clustering code is available from \url{https://gitlab.com/nwra/NullCluster}.

\facility{\textit{SDO}/HMI}


\bibliographystyle{aasjournal}

\bibliography{references}

\end{document}